\documentclass[letterpaper,aps,twocolumn]{revtex4}

\usepackage{amssymb}
\usepackage{amsmath}

\usepackage{graphics}      
\usepackage{graphicx}      
\usepackage{longtable}     
\usepackage{bm}            

\newcommand{\be}{\begin{equation}}
\newcommand{\ee}{\end{equation}}
\newcommand{\bqs}{\begin{equation*}}
\newcommand{\eqs}{\end{equation*}}

\begin{document}
\title{Pattern formation, traveling fronts and consensus versus fragmentation in a model of opinion dynamics}
\author         {Matt Holzer and Ratna Khatri}
\email          {mholzer@gmu.edu, rkhatri3@gmu.edu}
\affiliation {Department of Mathematical Sciences, George Mason University,  Fairfax, VA 22030, USA}
\date{\today}

\begin{abstract} We consider a continuous version of the Hegselmann-Krause model of opinion dynamics. Interaction between agents
either leads to a state of consensus, where agents 
converge to a single opinion as time evolves, or to a fragmented state with multiple opinions.  In this work, we linearize the system about a uniform density solution and predict consensus or fragmentation based on properties of the resulting dispersion relation.  This prediction is different depending on whether the initial agent distribution is uniform or nearly uniform.  In the uniform case, we observe traveling fronts in the agent based model and make predictions for the speed and pattern selected by this front.  

\end{abstract}

\maketitle

%

\section{Introduction} 
We study the dynamics of an agent based model of opinion dynamics described by the following system of ordinary differential equations, 
\be \label{eq:main}
 \frac{dx_i}{dt} = \sum_{j=1}^{N}\frac{\phi(|x_{j}-x_i|)}{\sum\limits_{k}\phi_{ik}}(x_{j}-x_{i})
\mbox{,} \hspace{0.5 cm} 
\ee
where there are $N$ total agents and $x_i(t)\in\mathbb{R}$ denotes the opinion of the $i$th agent and $\phi(x):\mathbb{R}\to\mathbb{R}$ with $\phi(x)\geq 0$ for all $x$.  This model is a continuous version of the Hegselmann-Krause model of opinion dynamics \cite{hegselmann02} and has been studied extensively in \cite{motsch14}.  Equation (\ref{eq:main}) describes the evolution of the opinion of the $i$th agent, which is influenced by the opinions of the remaining agents via a weighted sum.  The function $\phi:\mathbb{R}\to\mathbb{R}$ is called the interaction kernel and it ascribes the weights given to the opinions of other agents.  The key question we are concerned with here is how the properties of the interaction kernel affect the dynamics of the system, both in the long term and in the short term.

As in \cite{motsch14}, we will assume that the interaction kernel is compactly supported and therefore agents ignore those agents whose opinions are greater than some fixed distance from their own.  Without loss of generality we fix this distance to be one.  Note that if the interaction kernel is globally supported then all agents will eventually converge to a single opinion, see for example \cite{bertozzi09}.  When the interaction kernel is compactly supported the long term dynamics are more complicated and while the system always converges to an equilibrium state,  see \cite{jabin14}, it is not clear whether that state will constitute a single opinion; i.e. {\em consensus}, or whether multiple clusters with differing opinions will exist, i.e. {\em fragmentation}.

Whether the system will converge to a state of consensus or fragmentation depends on both the choice of interaction kernel and the initial conditions and is challenging to quantify in general.  In this article, we restrict our study to the case where the initial opinions of agents are distributed uniformly, or nearly uniformly, throughout some fixed domain and study how the choice of interaction kernel, the size of the initial domain and the density of the initial configuration affect whether the system converges to a state of consensus or fragmentation.  Our view is that this determination can often be made with only information concerning the linearization of the system about a homogeneous steady state and the pattern forming process that ensues.  We therefore focus on the pattern forming properties of (\ref{eq:main}) and  provide predictions for consensus or fragmentation based solely upon properties of the Fourier transform of $\phi(x)$.  

In Section~\ref{sec:pf}, we consider the case of nearly uniform initial conditions and predict that the pattern with the largest temporal growth rate in the linearized system will be observed, see also \cite{garnier17}.  In Section~\ref{sec:TF}, we contrast this with the case of uniform initial data.  Determining the inter-cluster separation in this case has attracted interest in the literature recently where it is sometime referred to as the $2R$ conjecture, see \cite{blondel07,blondel08,lorenz06,wang17}.  We interpret this process as a traveling front and use theory developed for fronts propagating into unstable states to derive a prediction for the inter-cluster distance, see \cite{vansaarloos03}.  This prediction matches well with numerical simulations and is distinct from the most unstable mode prediction.   Finally, in Section~\ref{sec:frag} we argue that on sufficiently large domains fragmentation is the norm.

We note that the field of opinion dynamics has received a great deal of interest in the literature over the past several decades, see for example \cite{bennaim03,blondel08,deffuant00,dittmer01,lorenz07,hegselmann02,hegselmann05,motsch14}.  
We also remark that while system (\ref{eq:main}) is motivated in terms of opinion dynamics, similar system also arise in problems related to bacterial aggregation, see \cite{bertozzi09,mogilner03,topaz06,vonbrecht12} among others.

\section{The mean field limit and dispersion relation}\label{sec:pf}
We find it convenient to use a mean field approximation.  Consider initial conditions for (\ref{eq:main}) consisting of a large number of agents distributed nearly uniformly throughout an interval.  Let $\rho = \rho(t,x)$ describe the density of agents with opinion $x$ at time $t$.  In the limit of a large number of agents, a mean-field evolution equation for $\rho(t,x)$ can be obtained which takes the form
\be
\label{eq:contmodel}
\partial _{t}\rho + \partial _{x}(\rho u)=0, \quad  u(x)=\frac{\int_{\mathbb{R}}^{}\phi(|y-x|)(y-x)\rho(y)dy}{\int_{\mathbb{R}}^{}\phi(|y-x|)\rho(y)dy},
\ee
see \cite{motsch14} for the derivation and discussion.

\subsection{The dispersion relation}
Since we are interested in initial conditions that are nearly uniformly distributed, it is natural to linearize (\ref{eq:contmodel}) about a constant density solution. Let $\rho(t,x) = 1 + \xi(t,x)$. The linearized system takes the form,
\be \label{eq:LinearSystem}
\xi_{t}=\frac{-\partial _{x}(\Phi*\xi)}{\int_{\mathbb{R}} \phi (x)dx},\quad 
\Phi(x)=\phi(x)x.
\ee
Consider solutions of the form
\begin{equation} \label{eq:SolForm}
\xi(t,x) = \xi_{o}e^{\lambda t+ikx}
\end{equation}
where, $ k$ denotes the wavenumber of the perturbation, and $ \lambda$ is the corresponding growth rate of the mode $k$.

Deriving solutions of the form (\ref{eq:SolForm}) leads us to a solvability condition known as the dispersion relation,
\begin{equation}
\lambda(k) = \frac{-ik\hat{\Phi}(k)}{\hat{\phi}(0)}, \label{eq:gendisp}
\end{equation}
where, $\hat{\Phi}(x)$ is the Fourier transform of $\phi(x)x$.

\begin{figure}[h]
\centering
\includegraphics[width=0.2\textwidth]{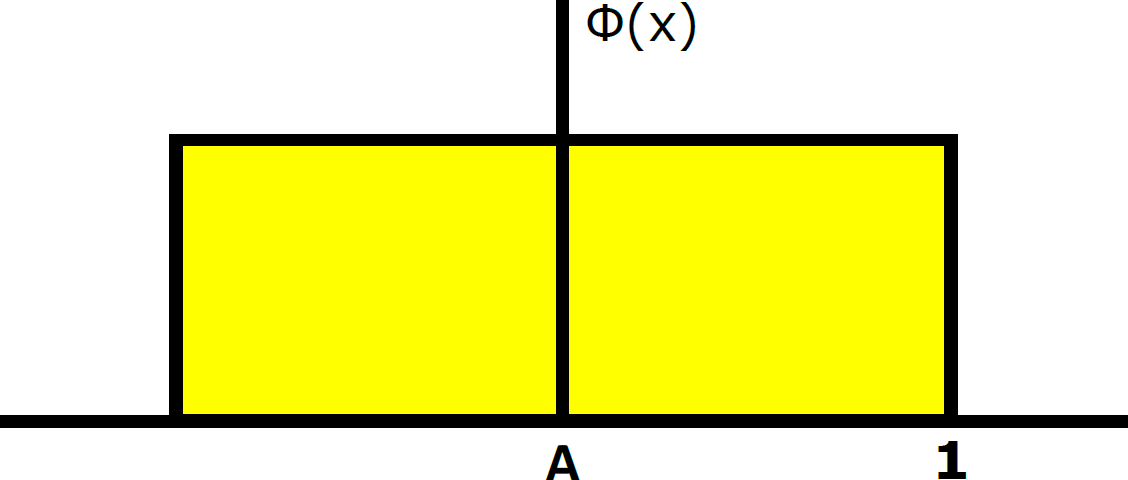} \hfil
\includegraphics[width=0.2\textwidth]{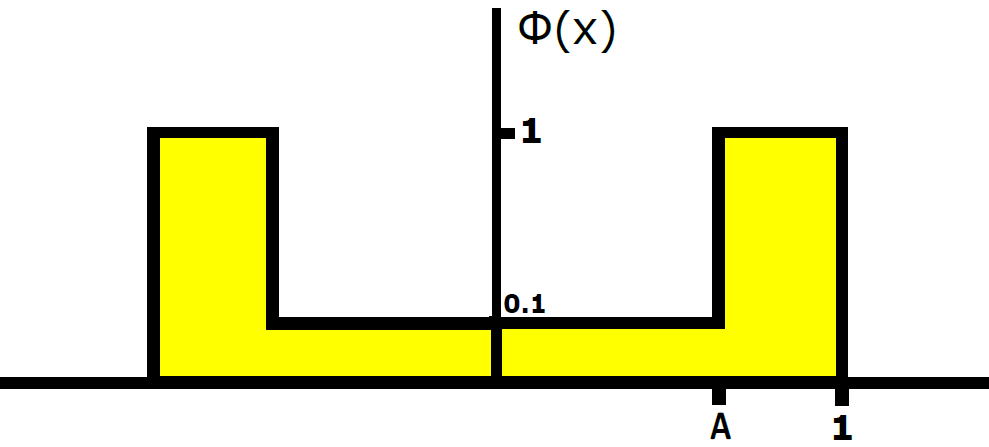}  
\caption{Examples of $\phi(x)$, where $\phi(x) \neq 0 \hspace{0.1 cm} \forall \hspace{0.1 cm} x \in [-1,1]$. The discontinuity in the kernel occurs at $\mathrm{A}$, where $\mathrm{A}=0$ in the figure on the \textit{left}, and $0<\mathrm{A}<1 $ in the figure on the  \textit{right}. \label{fig:kernels} }
\end{figure}

To focus the discussion we will consider two classes of kernels.  The first are those interaction kernels studied in \cite{motsch14} and depicted in Figure \ref{fig:kernels}, with definition
\begin{equation}\label{eq:kernel}
\phi_A(x)=\begin{cases}
\vspace{0.25 cm}
0 & |x| > 1\\   
\vspace{0.25 cm}
1 & A\leq |x|\leq 1 \\
0.1 & |x|< A \\
\end{cases} 
\end{equation}
The second class consists of exponential kernels,
\[ \phi_a(x)=e^{-a|x|}\chi_{[-1,1]}(x),\]
for $a\geq 0$.   

\begin{figure}[h]
\centering
   \includegraphics[width=0.2\textwidth]{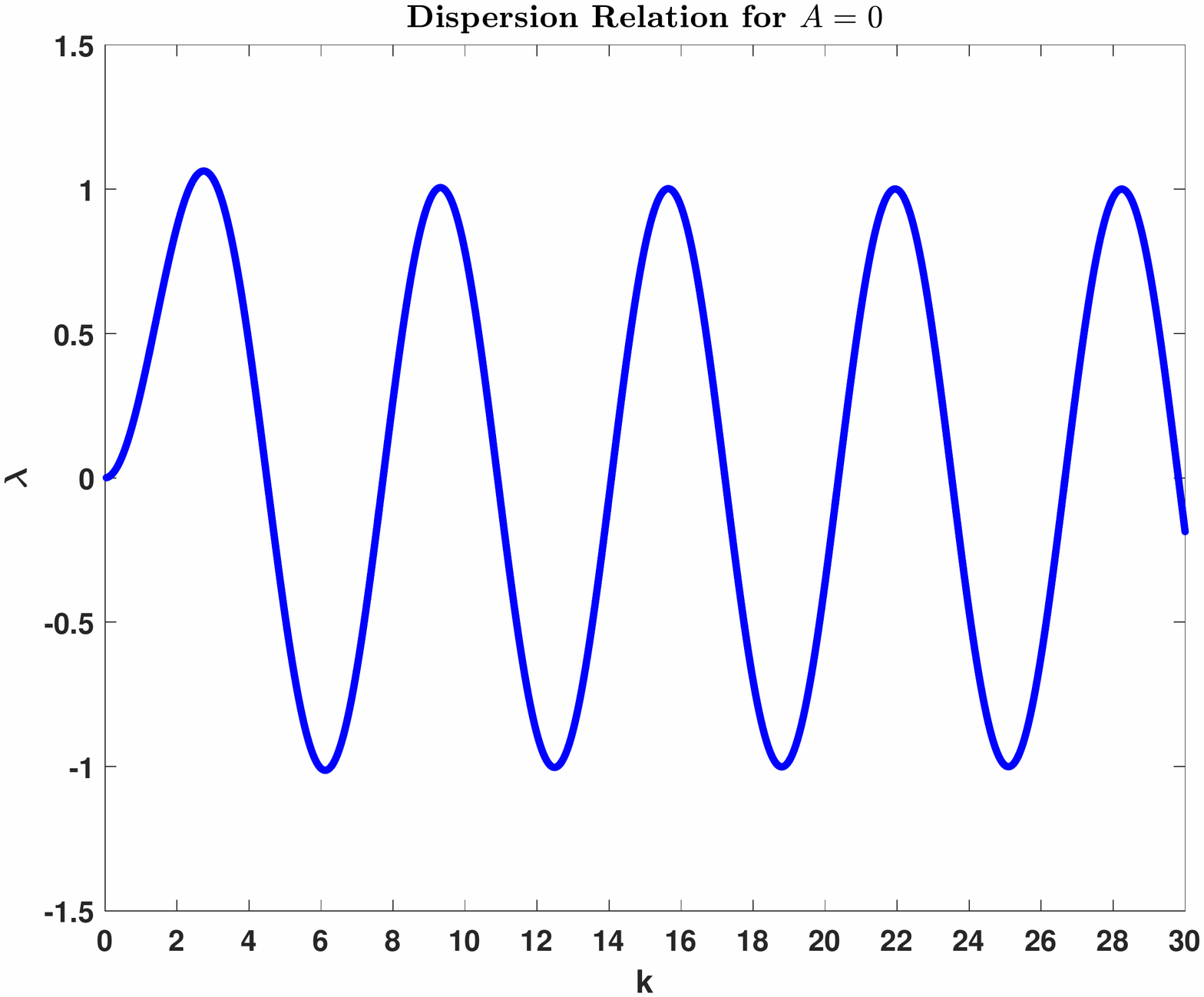} \hfil
   \includegraphics[width=0.2\textwidth]{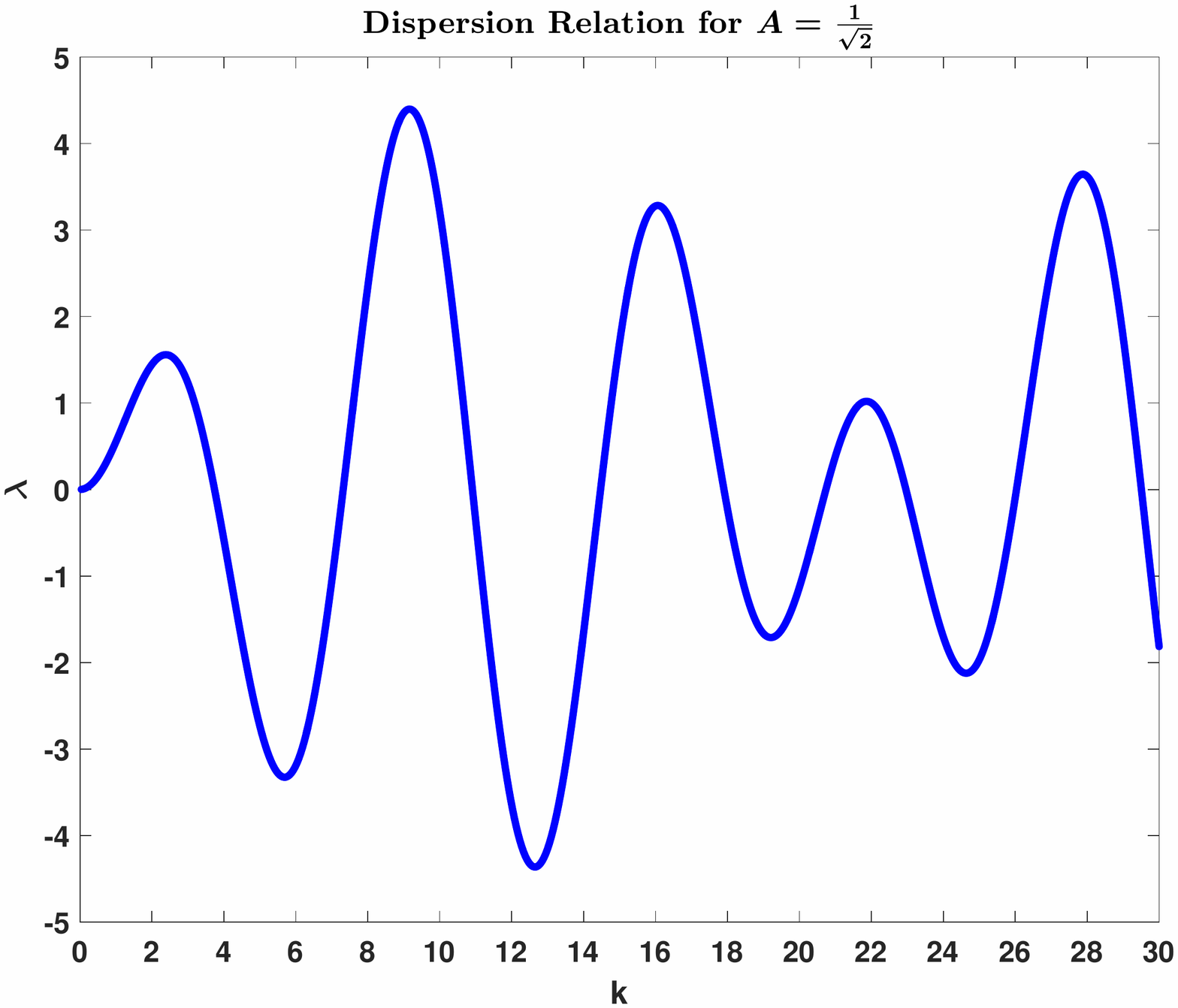} 
\caption{Dispersion relation of $\phi_A(x)$ for $\mathrm{A}=0$ \textit{(left)} and $\mathrm{A}=\frac{1}{\sqrt{2}}$ \textit{(right)}. Note that the period, $\frac{2\pi}{k_{max}}$, corresponding to the most unstable mode is $2.27$ and $0.68$, respectively.}
\label{fig:D}
\end{figure}

\begin{figure}[h]
\centering
 \includegraphics[width=0.2\textwidth]{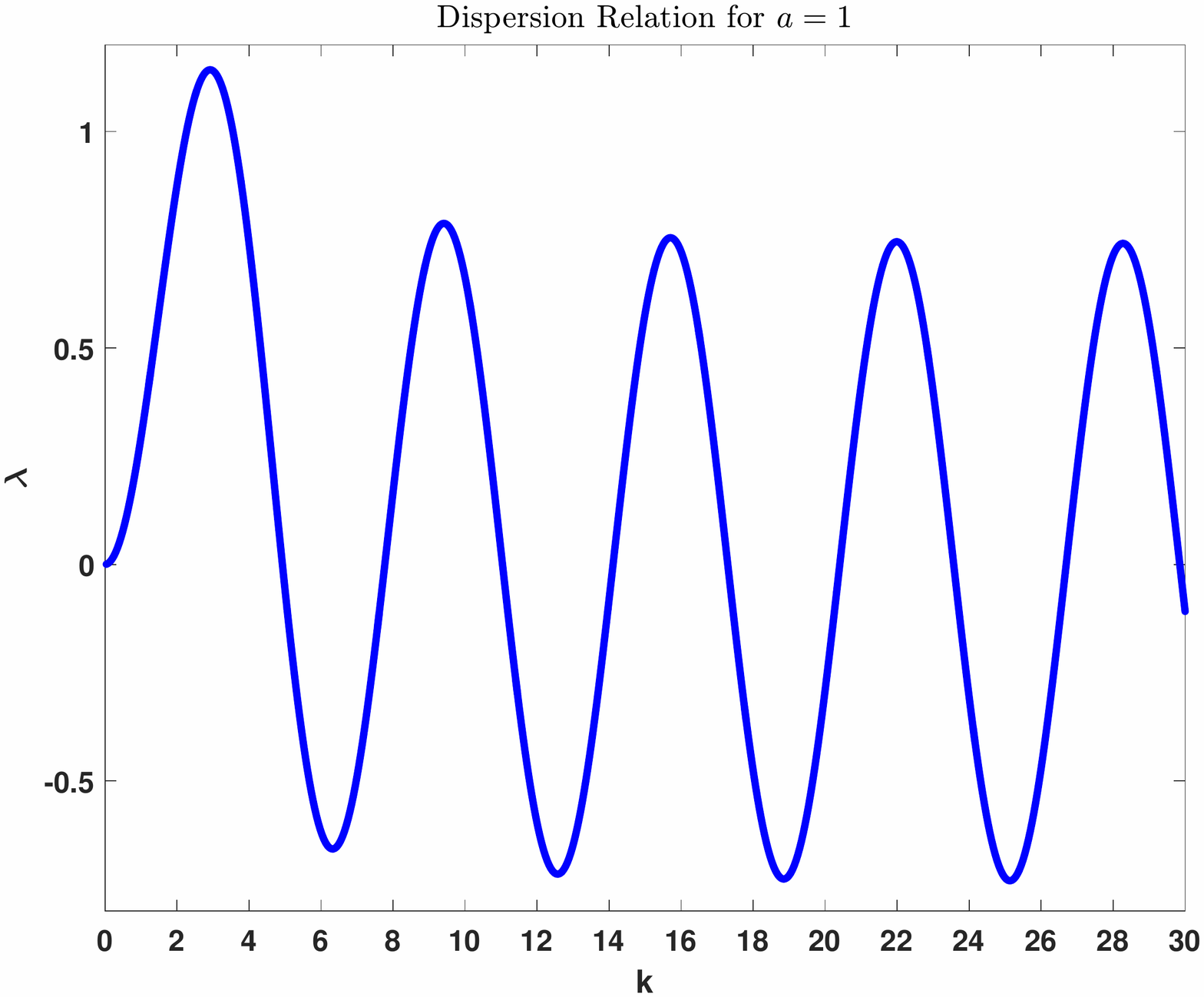} \hfil
   \includegraphics[width=0.2\textwidth]{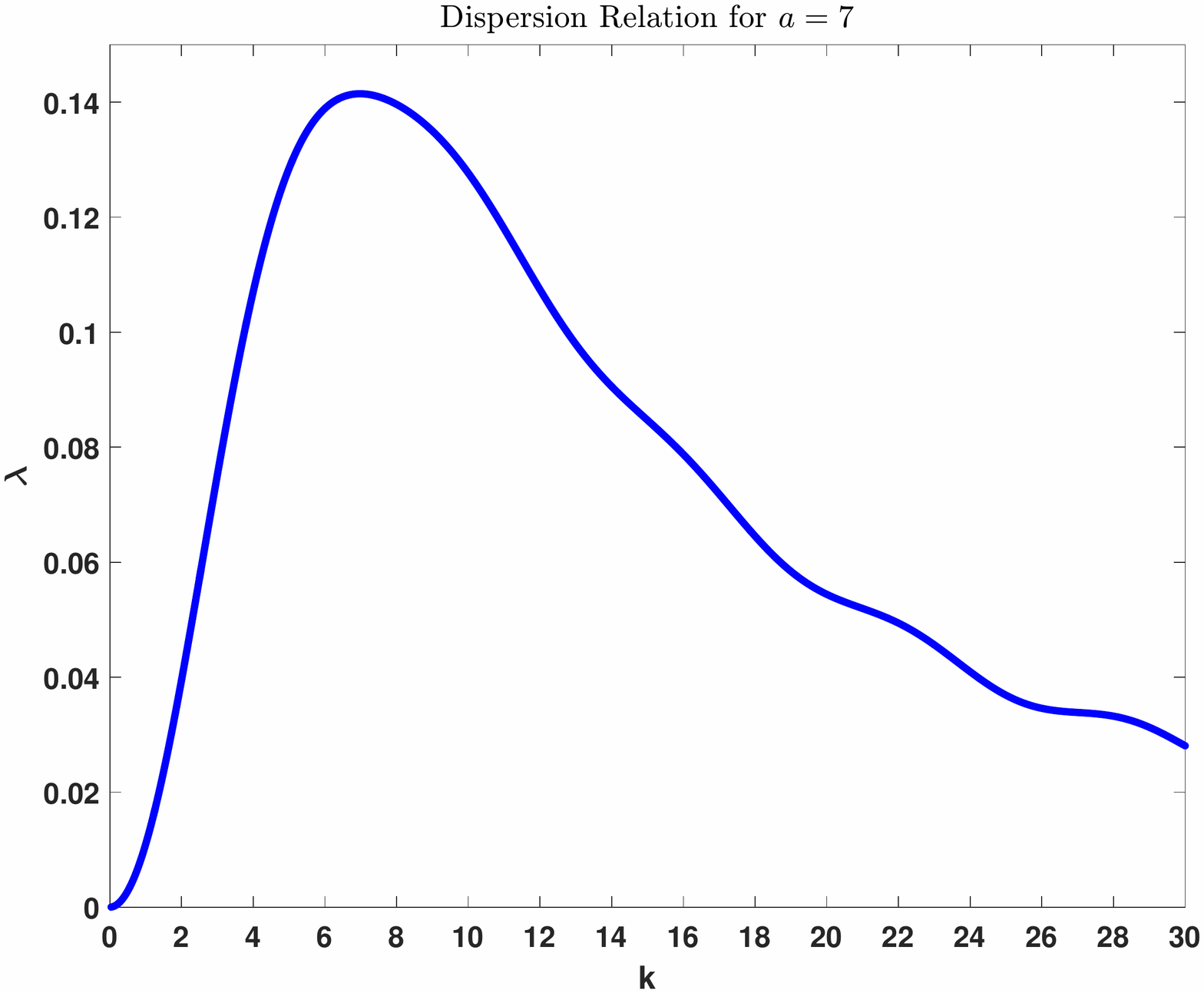} 
\caption{Dispersion relation of $\phi_a(x)$  for $a=1.0$ \textit{(left)} and $a=7.0$ \textit{(right)}.   Note that the period, $\frac{2\pi}{k_{max}}$, corresponding to the most unstable mode is $2.12$ and $1.16$, respectively. }
\label{fig:Da}
\end{figure}

The dispersion relation for both kernels can be computed explicitly.  For reference, we provide the dispersion relation for $\phi_A(x)$, 
\be
\lambda = \frac{1}{\hat{\phi}_A(0)} \left( \frac{9\mathrm{A}}{5}\cos k\mathrm{A}-2\cos k-\frac{9}{5}\frac{\sin k\mathrm{A}}{k}+2\frac{\sin k}{k} \right), \label{eq:disp}
\ee
with
\[
\hat{\phi}(0) = \frac{2}{10}\mathrm{A} + 2(1-\mathrm{A}).
\]
In Figure~\ref{fig:D} and Figure~\ref{fig:Da}, we plot the dispersion relations for $\phi_A(x)$ and $\phi_a(x)$ for representative values of $A$ and $a$, respectively.   Note that since the kernel is discontinuous the resulting dispersion relation does not decay as $k\to\infty$.  

\begin{figure}[h]
\centering
\includegraphics[width=0.2\textwidth]{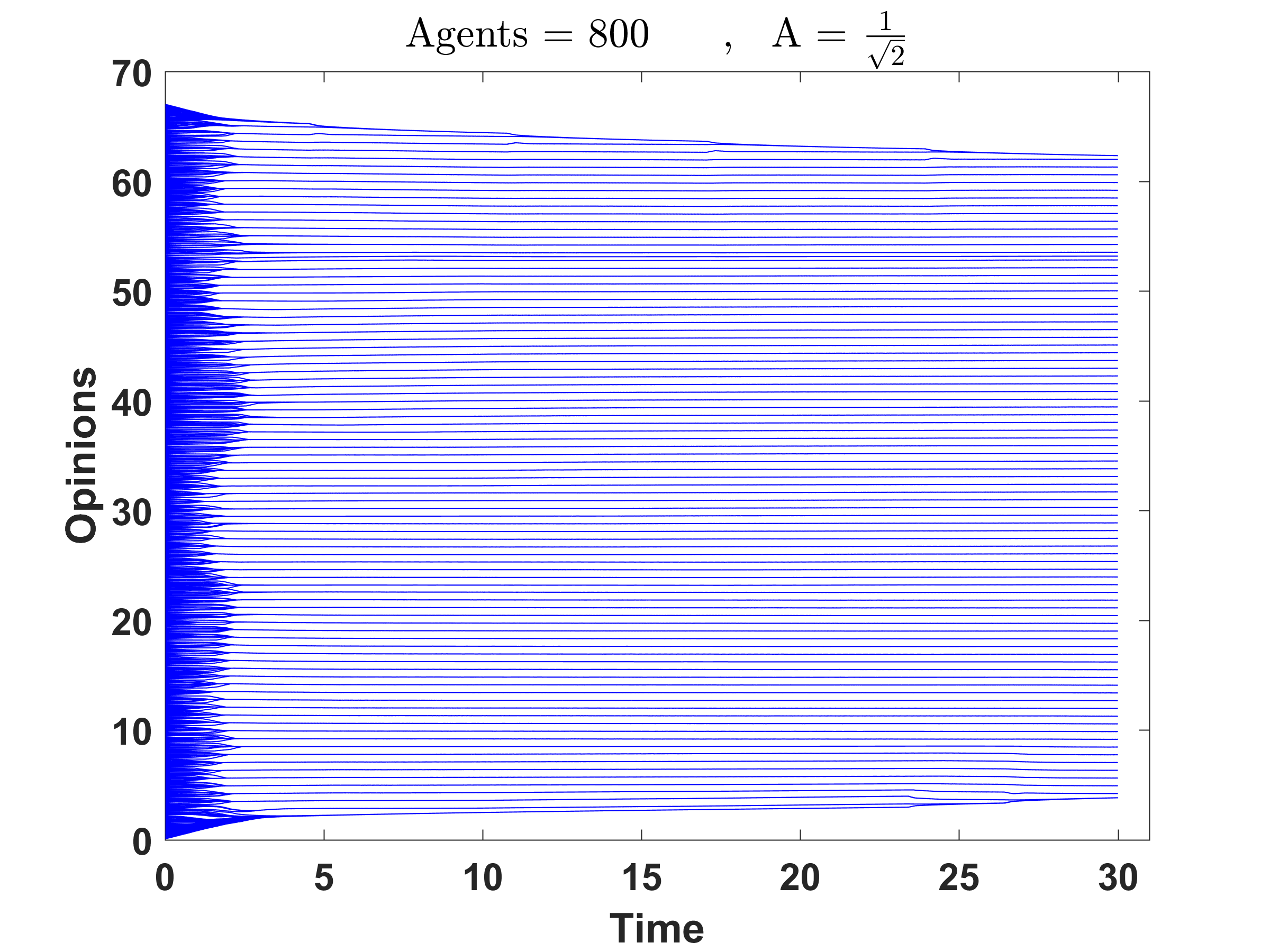} \hfil
\includegraphics[width=0.2\textwidth]{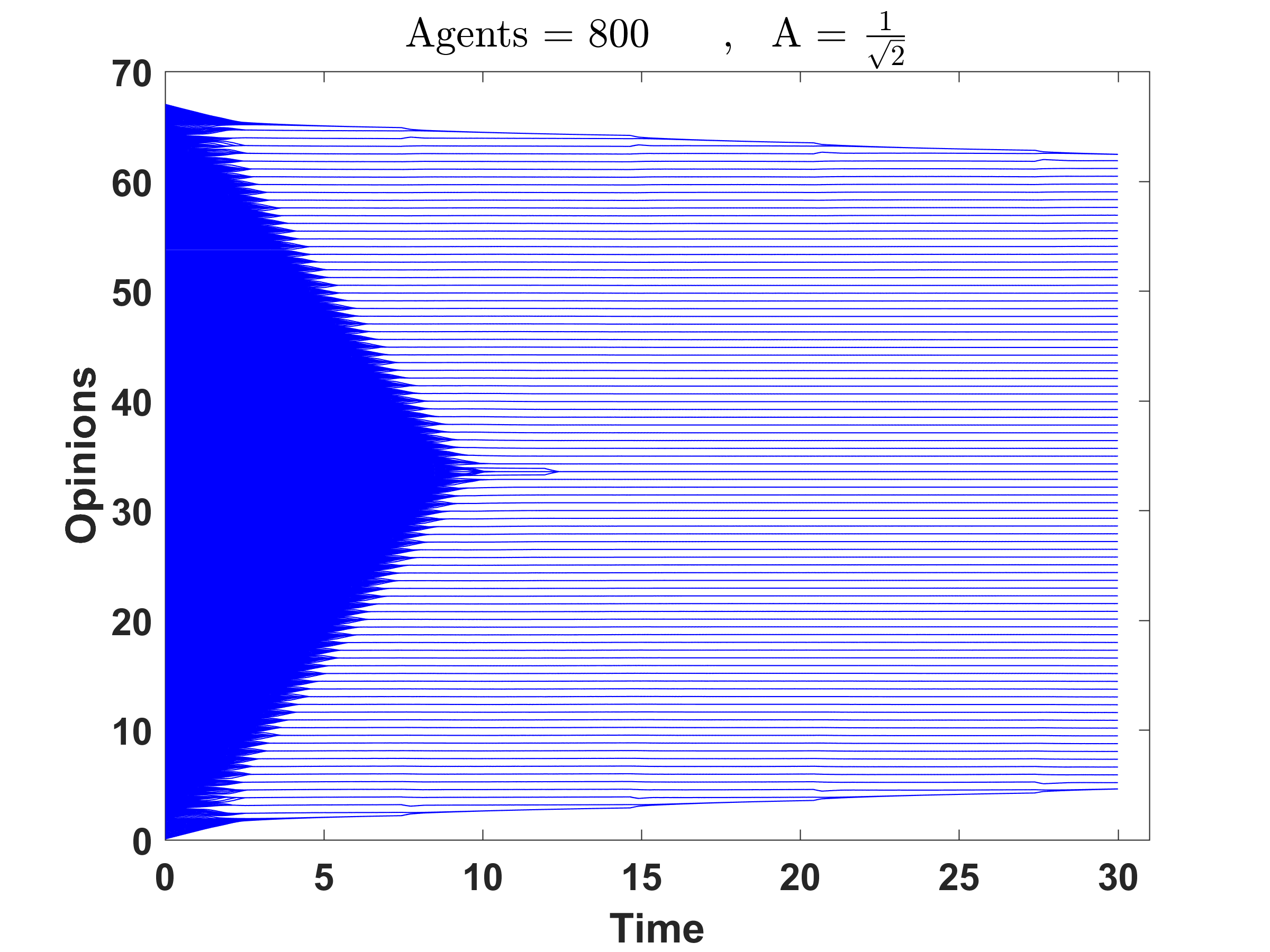}   
\caption{Numerical simulations of (\ref{eq:main}) with $A=\frac{1}{\sqrt{2}}$. Initial conditions are uniform \textit{(right)} and uniform with small, random perturbation \textit{(left)}.  In the uniform case, the agents aggregate into clusters following the passage of a traveling front.  In the case of a random perturbation, the system transitions immediately to a patterned state with approximately equal spacing. }\label{fig:simulationstwocases}
\end{figure}

We show the results of two numerical simulations in Figure~\ref{fig:simulationstwocases}.  We note the difference between the dynamics for uniformly distributed initial data and that of uniform initial data plus a small, random perturbation.  In the latter case, we observe an immediate aggregation of the agents into clusters.  In the former case, the interior agents are in equilibrium and the system transitions to its patterned state following the passage of a traveling front.  It is well known that the patterns selected through these two mechanisms do not have to be equal, \cite{kotzagiannidis12,vansaarloos03}.  In the remainder of this section we comment on the dynamics for random perturbations.  We will turn our attention to predicting the pattern determined by the invasion front in Section~\ref{sec:TF}.

\subsection{Predictions for nearly uniform initial conditions}

For the remainder of this section, we focus on initial conditions which are a small, random perturbation of a uniformly distributed initial condition.  The dispersion relation provides a relationship between the mode $e^{ikx}$ and its growth rate $\lambda$. For perturbations that excite all modes equally, we expect the most unstable modes to dominate the dynamics of the system. This has the effect of organizing agents into clusters with spacing approximately equal to  $\frac{2\pi}{k_{max}}$, where $k_{max}$ corresponds to the most unstable mode.   This leads to an immediate prediction of whether the system will converge to a state of consensus or fragmentation for nearly uniformly distributed initial conditions.  If the most unstable mode $k_{max}<2\pi$ then we expect the short time pattern formation process to push the system towards a state of fragmentation consisting of clusters of agents separated by approximately distance $\frac{2\pi}{k_{max}}>1$.  Conversely, if $k_{max}>2\pi$, we expect the system to be pushed towards a patterned state with period less than one.  The system remains connected in this scenario and we predict consensus.  

We refer the reader to \cite{garnier17} for an in depth discussion of validity of this prediction, for the class of kernels $\phi_A(x)$ and several other kernels.  We observe similar phenomena as noted in \cite{garnier17} and comment briefly on these observations.  If the dispersion relation predicts fragmentation then fragmentation is typically observed in simulations.  On the other hand, if the dispersion relation predicts consensus then the observed dynamic is more complicated.  For some values of $A$, for example $A\approx 0.7$, the system converges to consensus while for other values of $A$ the system will converge to consensus or fragmentation depending on the particular random perturbation performed.  

We expect, as in \cite{garnier17} that this uncertainty results from the fact that for some values of $A$ there are several modes with nearly identical temporal behavior and therefore a competition between these modes could lead to the selection of one or the other.  That being said, it is clear from the dispersion relations in Figure~\ref{fig:D} that for $A=0$ the temporal growth rates of the fragmentation and consensus modes are very close and yet fragmentation is always observed for this kernel.  Thus, it appears that the system prefers fragmentation through some mechanism that we do not fully understand.  Finally, we note that the above statements regarding consensus are true for sufficiently small sized domains.  If the domain is  made too large then fragmentation is observed, see Section~\ref{sec:frag}.

\section{Traveling fronts and  pattern selection}\label{sec:TF}
We now restrict ourselves to the case of initial opinion distributions that are uniformly distributed through a finite interval.  We assume that the number of agents is large compared to the domain size, so that the initial condition is dense and the mean field equations may be considered to approximate the dynamics.  For these  initial conditions, the agents in the center of the domain witness identical opinion distributions to the left and right and are in equilibrium.  In contrast, the agents on the extremes are not in equilibrium and adjust their opinions towards the center of mass.  This leads to the formation of traveling fronts
propagating from the edge of the domain towards the center and depositing in their wake clusters of agents, see also \cite{lorenz06}.  The goal of this section is to make predictions concerning the speed and pattern, or inter-cluster distance, selected by these fronts.  These quantities are of importance to the dynamics because they  determine the final state of the system and how quickly this state is attained.  

\subsection{Pattern selection}
We now focus on the pattern selected by the traveling invasion fronts.  The pattern generated by these fronts will determine whether the initial separation between groups of agents is larger than one, in which fragmentation occurs immediately, or if the separation is less that one, in which case fragmentation or consensus may occur depending on the size of the domain; see Section~\ref{sec:frag}.  There are two natural criterion for predicting  this selected pattern: the pattern corresponding to the most unstable mode of the dispersion relation and the pattern selected through invasion.  These are not, in general, equal.  It turns out that these two predictions are sufficiently close for the kernel $\phi_A(x)$ such that it is not possible to determine which mechanism is responsible for the selection of the pattern.  However, the kernel $\phi_a(x)$ does exhibit a significant deviation between these two predictions and we observe in numerical simulations that pattern is selected by the invasion front and is distinct from that of the most unstable mode.  We remark that determining the inter cluster distance for the kernel $\phi_A(x)$ with $A=0$ has been the focus of several articles and is sometimes referred to as the the $2R$ conjecture, see \cite{blondel07,blondel08,lorenz06,wang17}.  We argue that this separation is selected by properties of the traveling front.

\begin{figure}[h]
\centering
    \includegraphics[width=0.4\linewidth]{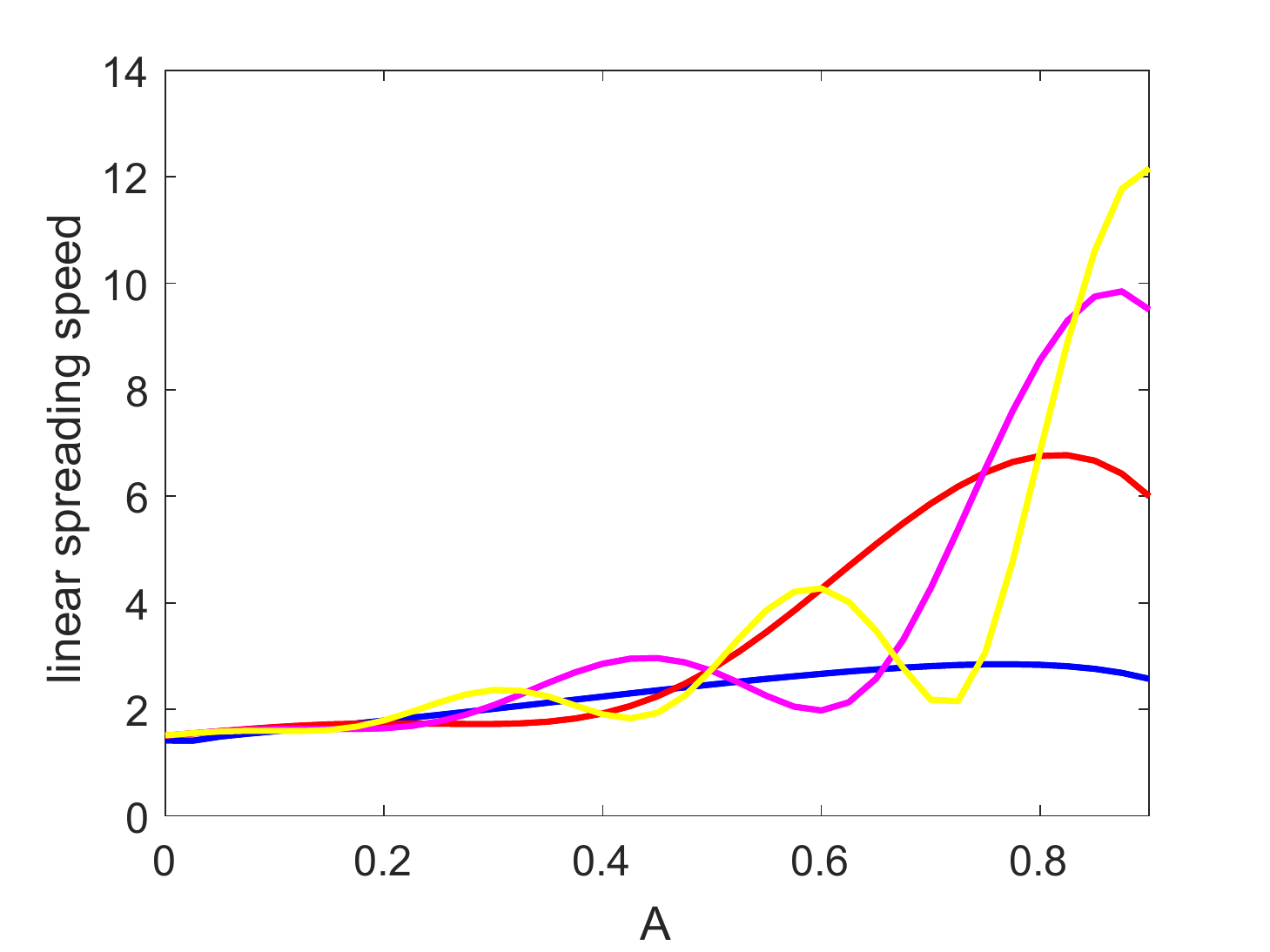}\hfil
    \includegraphics[width=0.4\linewidth]{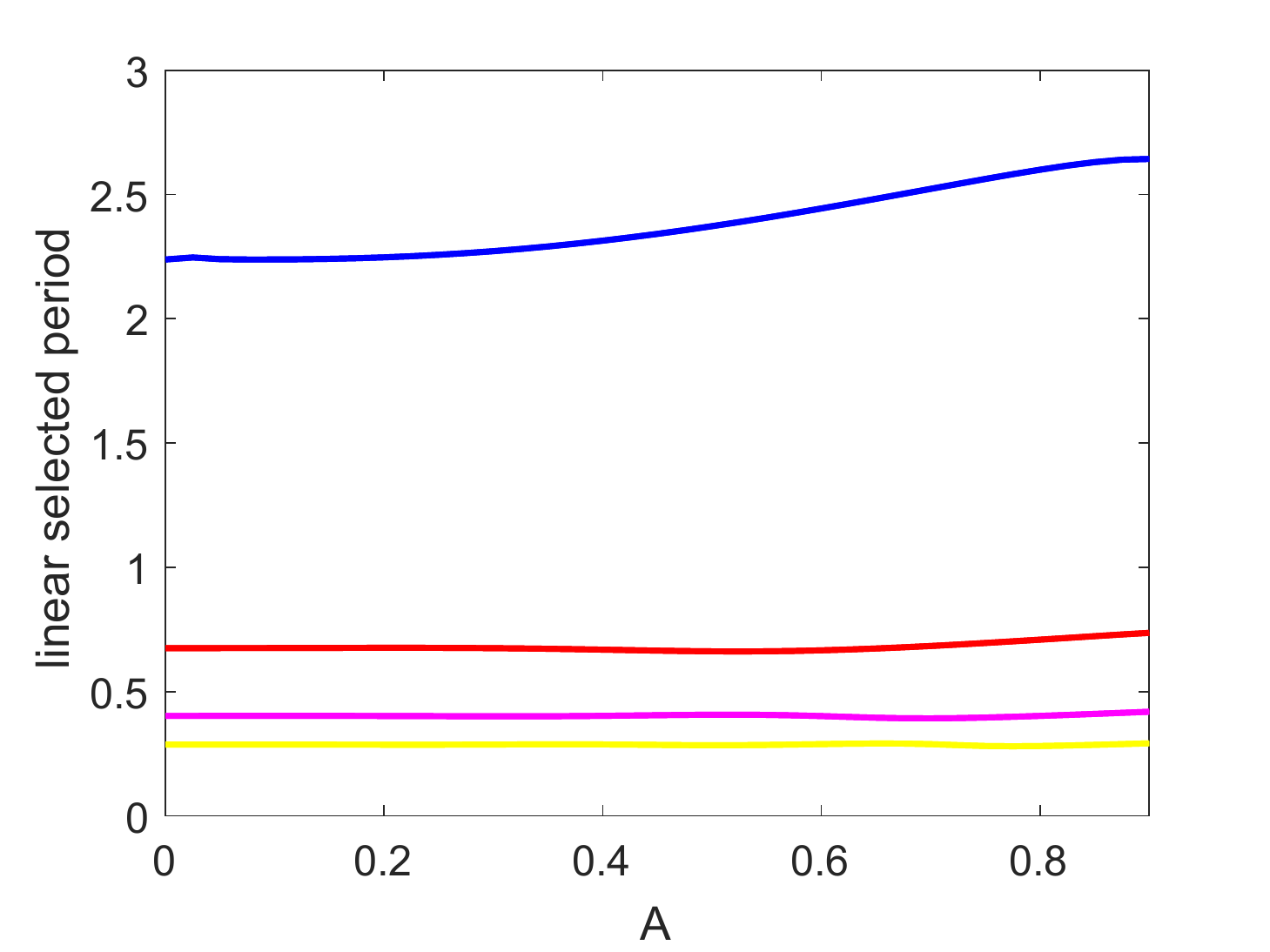}
\caption{Linear spreading speeds \textit{(left)} and selected patterns \textit{(right)} of the first four unstable modes.  The fragmentation mode is shown in blue, while the red, magenta and yellow curves depict consensus modes with increasing wavenumber, i.e. decreasing period. }\label{fig:spreadingspeed}
\end{figure}

In the mean field equations, the homogeneous steady state is linearly unstable and we can consider these fronts as front propagating into an unstable state.  An extensive amount of research has been conducted on front propagation into unstable states and we refer the reader to \cite{vansaarloos03} for a review.  Predictions for the invasion speed and pattern selected by the front can be determined from the linearization about the unstable state by locating pinched double roots of the dispersion relation in a co-moving frame, see \cite{holzer14,vansaarloos03}.  We consider the dispersion relation obtained as a solvability condition following the ansatz $\xi(t,x)=e^{\lambda t +\nu x}$ for $\lambda$ and $\nu\in\mathbb{C}$, wherein
\[ D_s(\lambda,\nu)=-\nu \frac{\int_{-1}^1 z\phi(z)e^{\nu z}dz }{\int_{-1}^1 \phi(z)dz} +s\nu-\lambda.\]
Taking $\nu=ik$ and $s=0$ recovers the equation (\ref{eq:gendisp}), however, in a moving coordinate frame the relevant modes are fully complex and it is more natural to consider $\nu\in\mathbb{C}$.  
The {\em linear spreading speed} is determined by locating solutions $s_{lin}\in\mathbb{R}$, $\omega\in\mathbb{R}$, and $\nu\in\mathbb{C}$ solving the system of complex equations
\be D_{s_{lin}}(i\omega_{lin},\nu)=0,\quad \partial_\nu D_{s_{lin}}(i\omega_{lin},\nu)=0.\label{eq:PDR} \ee
In a frame moving at the linear spreading speed the front is oscillating in time with frequency $\omega_{lin}$.  Thus, the {\em linearly selected wavenumber} and {\em linearly selected period} are
\[ k_{lin}=\frac{\omega_{lin}}{s_{lin}}, \quad L_{lin}=\frac{2\pi s_{lin}}{\omega_{lin}}.\] 
The dispersion relation for the kernel $\phi_A(x)$ is
\begin{eqnarray} 
 D_s(\lambda, \nu)&=& 
 \frac{2}{\hat{\phi}(0)} \left( \frac{-9\mathrm{A}}{10}\cosh \nu \mathrm{A}+\cosh \nu+\frac{9}{10}\frac{\sinh \nu\mathrm{A}}{\nu} \right. \nonumber \\
&-&\left.\frac{\sinh \nu}{\nu} \right)+s\nu-\lambda, \label{eq:D}
 \end{eqnarray}
For varying values of $A$, we compute predictions for the linear spreading speed and linearly  selected pattern of the first four modes; see Figure~\ref{fig:spreadingspeed}.  We note that local maximums of the dispersion relation (\ref{eq:disp}) are pinched double roots with $\mathrm{Re}(\lambda)>0$ and these roots can be numerically continued in the parameter $s$ until the condition (\ref{eq:PDR}) is satisfied.

\begin{figure}[h]
\centering
\includegraphics[width=0.2\textwidth]{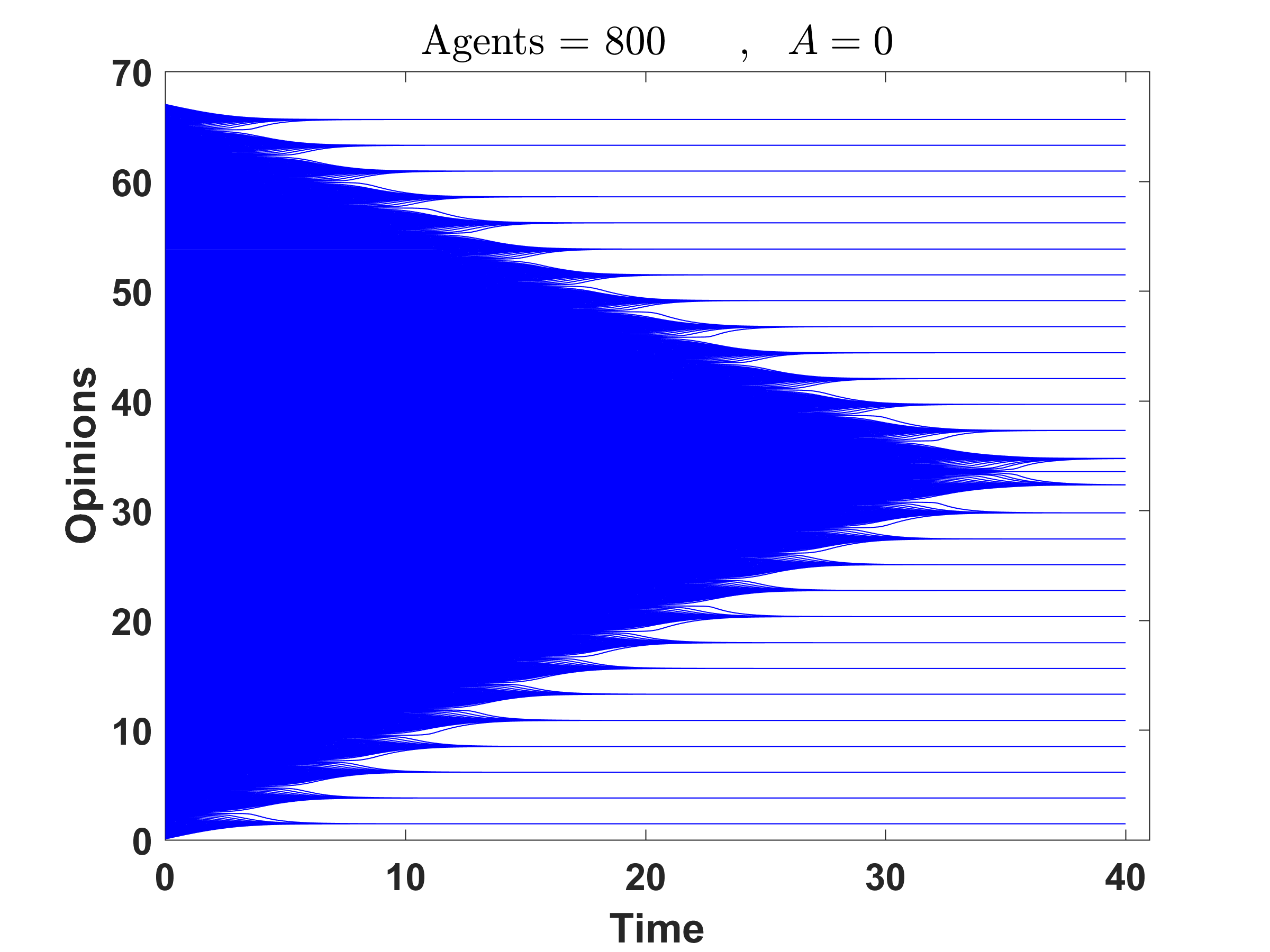} \hfil
\includegraphics[width=0.2\textwidth]{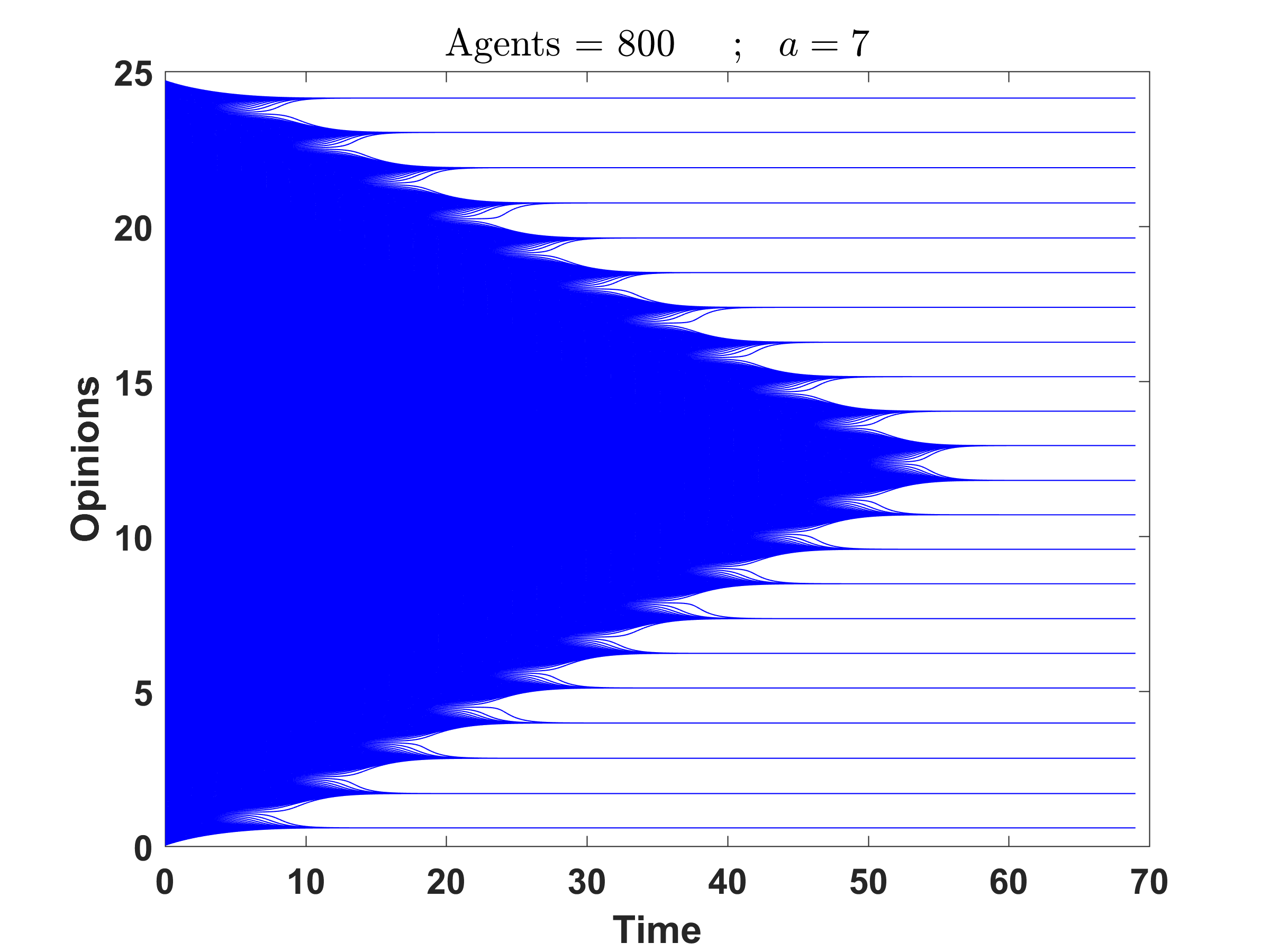}   \\
\caption{Fronts in two simulations of (\ref{eq:main}).  The case of $\phi_A(x)$ for $A=0.0$   \textit{(left)} and  $\phi_a(x)$ for $a=7.0$  \textit{(right)}.  Note that the front for $a=7.0$ selects a pattern with period $L_{lin}>\frac{2\pi}{k_{max}}$, recall Figure~\ref{fig:Da}.   }\label{fig:simulations}
\end{figure}
We now comment on our observations.  As stated above, we do not find a significant difference between $k_{max}$ and $k_{lin}$ for the kernel $\phi_A(x)$.  This feature makes it difficult to distinguish between the possible mechanisms leading to the selection of the inter cluster distance and numerical simulations reveal close agreement between this prediction and observed inter cluster distances. 

\begin{figure}[h]
\centering
    \includegraphics[width=0.6\linewidth]{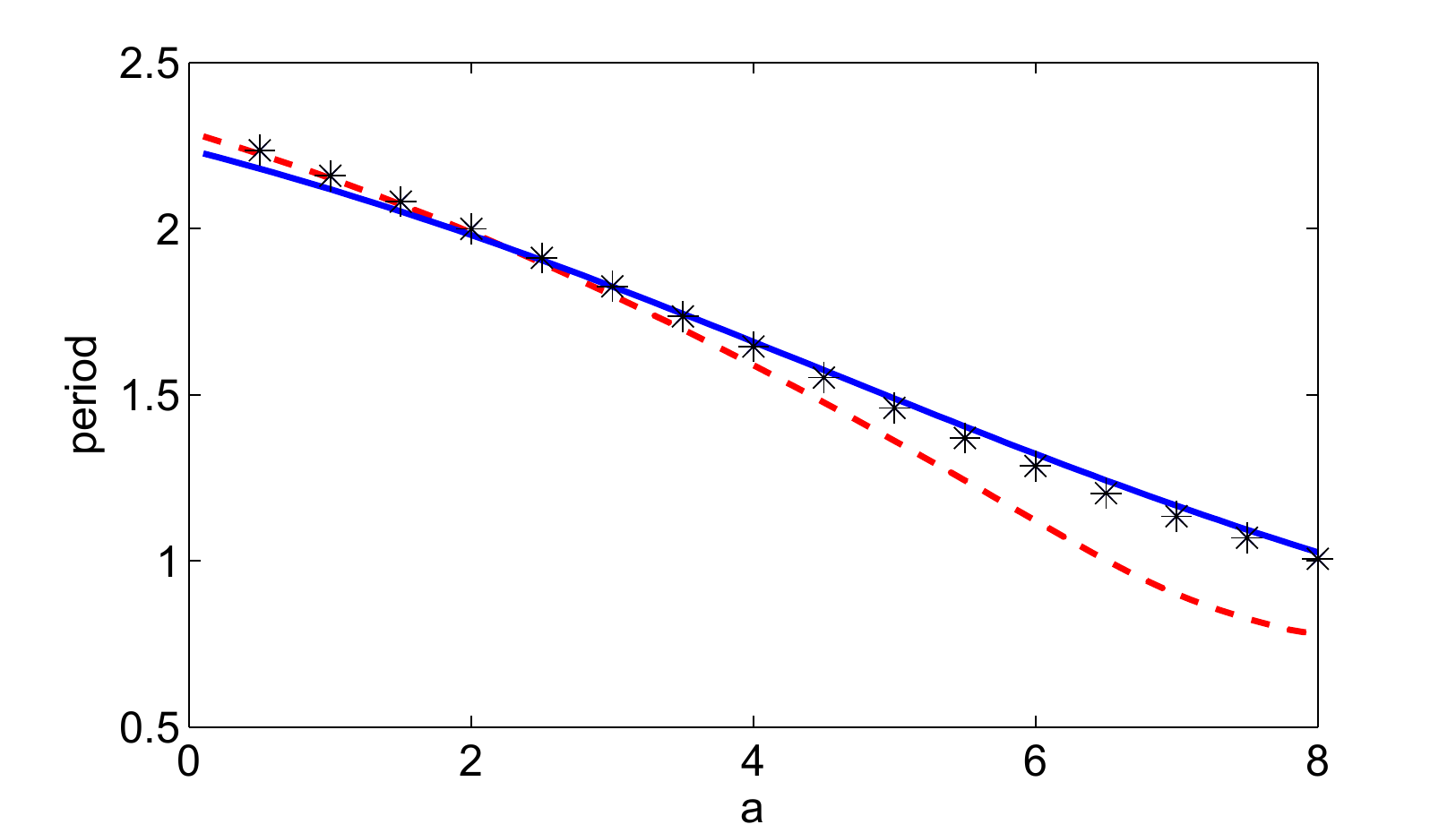}\hfil
\caption{Linearly predicted periods for the kernel $\phi_a(x)$ for $0<a\leq 8$.  The blue solid line is the prediction based upon the invasion process, $\frac{2\pi}{k_{lin}}$ while the red dashed line is the prediction based upon the most unstable mode of the dispersion relation, $\frac{2\pi}{k_{max}}$. The asterisks denote the pattern observed in numerical simulations -- simulation parameters are $600$ agents on the domain $[0,24.7]$, with agents inside $[0,1]$ fixed for all time.  The simulations were performed using explicit Euler with timestep of $\delta t=0.01$.    }\label{fig:expkernel}
\end{figure}

In contrast, the kernel $\phi_a(x)$ does reveal noticeable differences between $k_{max}$ and $k_{lin}$.  These predictions are plotted in Figure~\ref{fig:expkernel} for $0<a<8$ together with observed periods from numerical simulation.  We comment on several interesting features.  For small values of $a$ these wavenumbers are in close agreement.  However, above 
$a\approx 3$, these predictions deviate from one another.  In fact, for $a$ sufficiently large the most unstable mode $k_{max}$ predicts consensus while the linearly selected mode predicts fragmentation.  Numerical simulations reveal selected patterns that closely agree with the linearly selected mode, see Figure~\ref{fig:simulations} and  Figure~\ref{fig:expkernel}.

\subsection{Speed selection}
The speed of the front is an important quantity as it determines the timescale on which the original pattern forming process occurs.  When the front selects a pattern with period greater than one then the quantity $\frac{L}{2s_{lin}}$ provides a prediction for how long it takes the system to converge to equilibrium.   We now compare speed predictions derived from the pinched double root criterion to those observed in numerical simulations.  We again emphasize that the simulations are performed for the agent based model (\ref{eq:main}) while the predictions are based upon the dispersion relation derived from the mean field approximation (\ref{eq:contmodel}).  In the simulations of (\ref{eq:main}) the location of the front is said to be the position in space where the rate of change of the opinion of a certain agent first exceeds a small threshold.  Comparing this position at two different times we find an estimate for the speed.  To reduce computational time and allow us to consider systems with larger densities we make two simplifications.  First, we artificially fix the opinions of the extreme agents at one end of the domain and effectively consider the problem on half of the original domain.  Second, we use a technique of \cite{bennaim15} and periodically shift the domain -- forgetting some agents behind the front and adding new agents with the same dense uniform distribution ahead of the front.

In Table I we focus on two cases $A=0.0$ and $A= 0.7$  and compile numerically observed spreading speeds for (\ref{eq:main}) and predictions based upon the double root criterion applied to (\ref{eq:D}).  For the case of $A=0.0$, the fragmentation mode dominates and the observed spreading speeds are increasing functions of the initial agent density.  The mismatch between observed and selected speeds is likely due to the slow convergence of spreading speeds between mean field equations and their finite population analogs, see for example \cite{brunet97}.  In the agent based model, the fronts are propagating into a state that is stable against sufficiently small perturbations and this slows the front propagation.  However, we note that even with very high density initial conditions there remains a significant gap between the observed spreading speeds and those predicted by the mean field approximation.  The speeds for larger $A$ also remain considerably below the mean field approximation and do not vary monotonically as a function of initial density.    It is remarkable that the inter cluster distances are so well approximated by $L_{lin}$ in spite of the poor match in speed.  It would be interesting to understand this relationship in more detail.  

\begin{table}
  \begin{tabular}{ | c | c  | c |}
    \hline
    Agent separation & $A=0.0$  & $A=0.7$  \\ \hline
   $0.098$  & $0.974$   & $4.350$\\ \hline
   $0.049$ & $1.039$ & $4.179$\\ \hline
    $0.024$ & $1.100$ &$4.270$ \\ \hline
   $0.012$ & $1.202$ & $4.876$ \\ \hline
  mean field & $1.414$ & $5.832$   \\
    \hline
  \end{tabular}
\caption{ Observed spreading speeds in numerical simulations for varying initial agent separations and their comparison to the mean-field spreading speed.  We emphasize that these represent the observed spreading speeds in one numerical trial and will vary as the threshold and time of simulation is changed.  However, the overall trends are preserved as across various simulation parameters.  Also, note that the mean field speed given in the table for $A=0.0$ is the speed of the fragmentation mode.  The actually spreading speed is approximately $1.502$ and corresponds to a consensus mode.     }
\end{table}

\section{Fragmentation for consensus kernels on  large domains}\label{sec:frag}
We now comment on the dynamics following the initial pattern forming process.  Following this initial transient we expect that the system will have organized itself into clusters of agents with nearly identical opinions.  If the inter cluster distance exceeds one then the system is near equilibrium and the configuration will not change.  Alternatively, if the inter cluster distance is smaller than one then we expect the possibility of consensus as the extreme agents slowly moderate their opinions.  Indeed, on small to medium sized domains this is what is observed and the system transitions to a state of consensus.   However, on sufficiently large domains we argue that the system will instead converge to a state of fragmentation as the extreme groups become so large as to rip the pattern apart.

\begin{figure}[h]
\centering
    \includegraphics[width=0.4\linewidth]{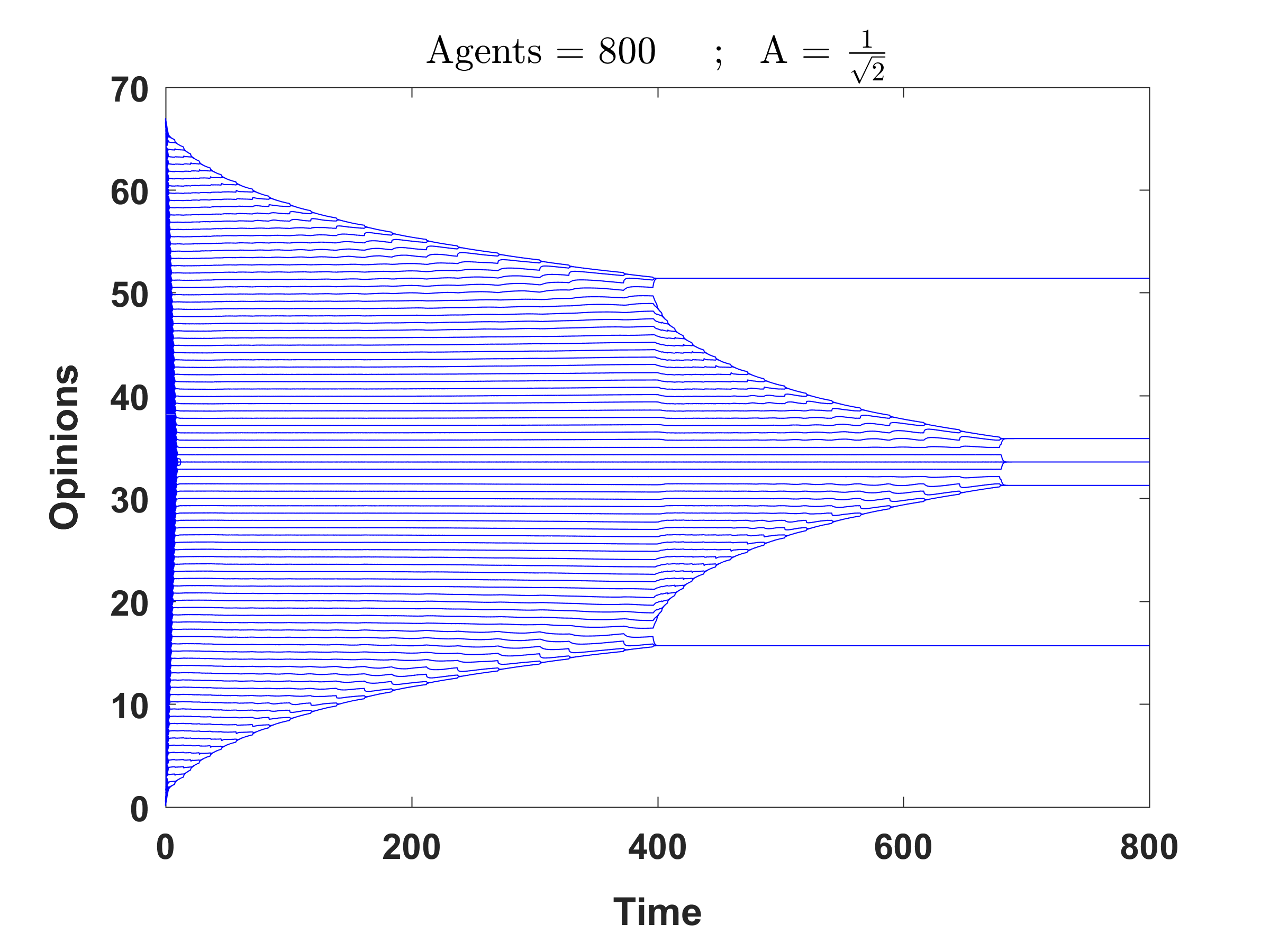}\hfil
    \includegraphics[width=0.4\linewidth]{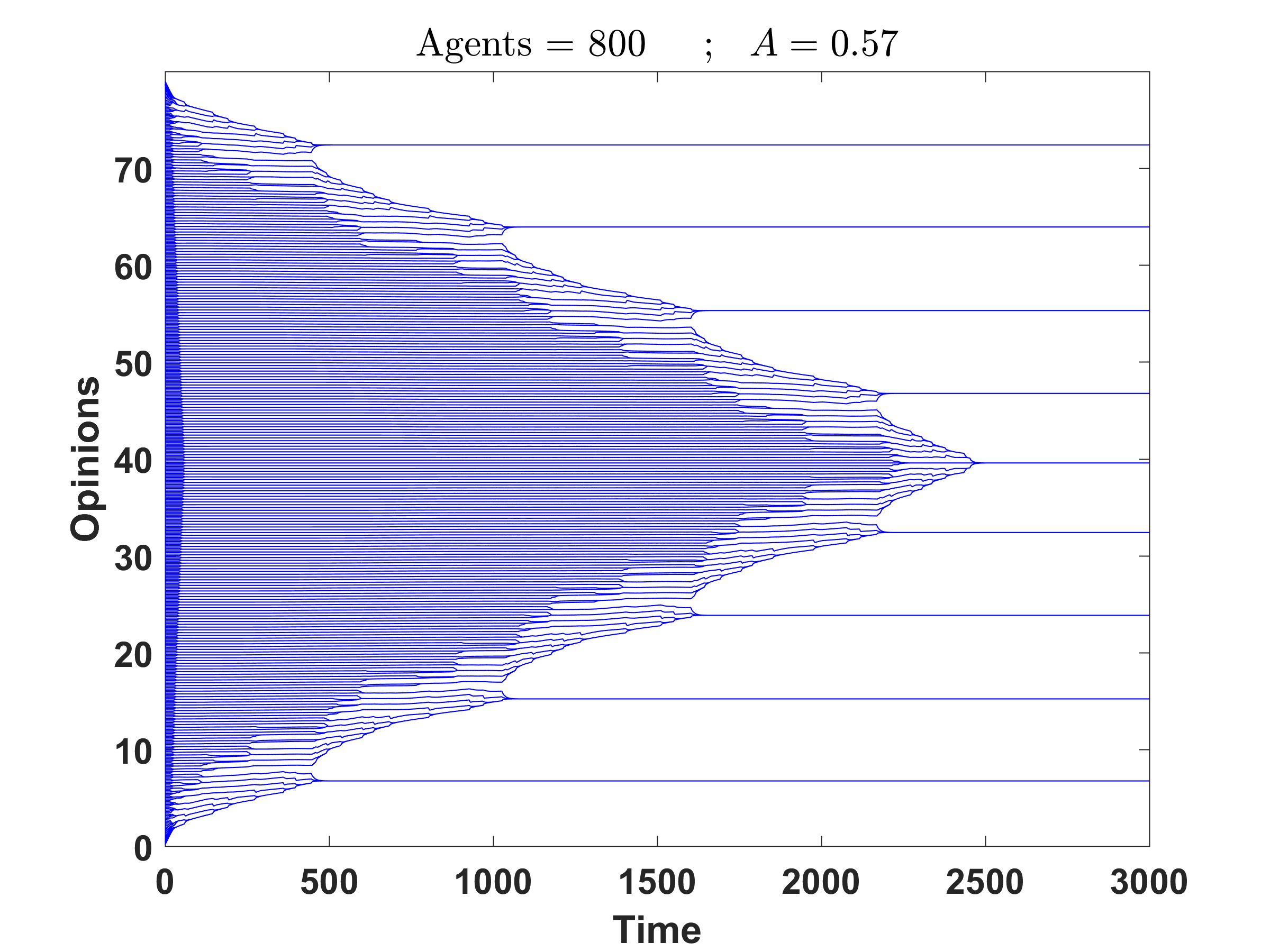}\hfil
\caption{Numerical simulations of (\ref{eq:main}) for $A=\frac{1}{\sqrt{2}}$ \textit{(left)} and $A=0.57$ \textit{(right)}.  Note the difference in time scale from Figure~\ref{fig:simulations}.  Following the passage of the traveling front, the system is configured into a patterned state.  The extreme group of agents slowly converges towards the center, until eventually the pattern is ripped apart and fragmentation occurs. Note the existence of a secondary coarsening process that occurs in the case of $A=0.57$.  }\label{fig:fragforlargeL}
\end{figure}

This is borne out in numerical simulations, see Figure~\ref{fig:fragforlargeL}, and we briefly comment on the mathematical justification of this phenomena.  Following an initial transient, we assume that the system has arranged itself into clusters of agents with opinions $ x_1<x_2<x_3 \dots <x_m.$
By considering a representative agent from each cluster the dimension of the system is dramatically reduced.  We assume that the system has evolved for some time and take the ratio of the size of the extreme groups to the interior groups as a parameter, $M$, and suppose for simplicity that the spacing between groups is greater than one half, but less than one.  In this case, the dynamics of (\ref{eq:main}) is reduced to
\begin{eqnarray*}
\frac{dx_1}{dt}&=& \frac{\phi(x_2-x_1)}{M\phi(0)+\phi(x_2-x_1)}(x_2-x_1) \\
\frac{dx_2}{dt}&=& \frac{M\phi(x_2-x_1)}{\phi(0)+M\phi(x_2-x_1)+\phi(x_3-x_2)}(x_1-x_2) \\
&+&  \frac{\phi(x_3-x_2)}{\phi(0)+M\phi(x_2-x_1)+\phi(x_3-x_2)}(x_3-x_2)\\
\frac{dx_3}{dt}&=&  \frac{\phi(x_3-x_2)}{\phi(0)+\phi(x_3-x_2)+\phi(x_4-x_3)}(x_2-x_3) \\
&+&  \frac{\phi(x_4-x_3)}{\phi(0)+\phi(x_3-x_2)+\phi(x_4-x_3)}(x_4-x_3)\\
 &\vdots & 
\end{eqnarray*}
In the limit $M\to\infty$, we find that $\frac{dx_1}{dt}=0$ while $\frac{dx_2}{dt}=(x_1-x_2)$.  Thus, we expect the extreme cluster to remain fixed to leading order while the secondary cluster will converge exponentially in time to the extreme cluster.  In this way, we expect fragmentation to occur for sufficiently large initial opinion distributions -- even when the interaction kernel predicts consensus, see Figure~\ref{fig:fragforlargeL}.

\section*{Acknowledgments}
\vspace{-.4cm}
This research was conducted as part of the EXTREEMS program at George Mason University and RK was partially funded by the NSF (DMS-1407087).  MH was partially supported by the NSF (DMS-1516155).  The authors thank Tyrus Berry for assistance with the numerical simulations.

\bibliographystyle{abbrv}
\bibliography{TWODmaster}

\end{document}